%% file: main.tex
\documentclass[a4paper, amsfonts, amssymb, amsmath, reprint, showkeys, nofootinbib, twoside]{revtex4-1}
\usepackage[english]{babel}
\usepackage[utf8]{inputenc}
\usepackage[colorinlistoftodos, color=green!40, prependcaption]{todonotes}
\input{preamble}
\usepackage[pdftex, pdftitle={Article}, pdfauthor={Author}]{hyperref} 
\begin{document}
\title{Stochastic systems with Bose-Hubbard interactions:\\ Effects of bias on particles {on a} $\mathbf{1}$D lattice }

\author{Swastik Majumder${}^{1,2}$}
\email{majumderswastik09@gmail.com}

\author{Mustansir Barma${}^{1}$}
\email{barma@tifrh.res.in}
\affiliation{${}^{1}$ Tata Institute of Fundamental Research, Hyderabad,\\ 36/P, Gopanpally Village, Serilingampally Mandal, Hyderabad, Telangana 500046}
\affiliation{${}^{2}$ Indian Institute of Science Education and Research Kolkata,\\ Campus Rd, Mohanpur, Haringhata Farm, West Bengal 741246, India}
\date{\today} 

\begin{abstract}
Driven non-equilibrium lattice models have wide-ranging applications in contexts such as mass
transport, traffic flow, and transport in biological systems. In this work, we investigate the steady-state properties of a one-dimensional lattice system that allows multiple particle
occupancy {on each} site. The particles undergo stochastic nearest-neighbor jumps influenced by both a
directional bias and on-site repulsive interactions {of the Bose-Hubbard type}. With periodic boundary conditions, we observe a
non-monotonic dependence of inter-site correlation functions on the interaction strength. At large
interaction strengths, {the state consists of quiescent stacks of stationary particles along with an emergent asymmetric simple exclusion process(ASEP)}, and the particle current exhibits a periodic dependence on density. In contrast, with open boundary conditions, the
system displays step-like density profiles reminiscent of those in tilted Bose-Hubbard systems, {and a regime with}
a macroscopic number of empty sites followed by a steep parameter-dependent increase in density.
Our results highlight how the interplay between drive, interaction,
and boundary conditions {leads to distinctive signatures on the current and density profiles in the steady state in different regimes.} 
\end{abstract}

\maketitle

\input{sections/section01.tex} 
\input{sections/section02.tex}
\input{sections/section03.tex}
\input{sections/section04.tex}
\input{sections/section06.tex}
\input{sections/acknowledgements.tex}

\end{document}

%% file: preamble.tex
\usepackage{amsthm}
\usepackage{mathtools}
\usepackage{caption}
\usepackage{ragged2e}  

\usepackage[justification=justified, font=small, labelfont=bf]{caption}
\usepackage{physics}
\usepackage{booktabs}
\usepackage{siunitx}
\usepackage{xcolor}
\usetikzlibrary{decorations.pathreplacing}
\usepackage{graphicx}
\usepackage[left=23mm,right=13mm,top=35mm,columnsep=15pt]{geometry} 
\usepackage{adjustbox}
\usepackage{placeins}
\usepackage[T1]{fontenc}
\usepackage{lipsum}
\usepackage{csquotes}
 \usepackage{ragged2e}
\usepackage{caption}
\usepackage{subcaption}
\usetikzlibrary{arrows.meta}
\usepackage[justification=justified,
             singlelinecheck=false,
             labelfont=normalfont,
             labelsep=period,
             font=small]{caption}

\usepackage{titlesec}

%% file: sections/section01.tex
\section{Introduction} \label{sec:Introduction}

This paper {is concerned with} a class of driven, non-equilibrium lattice models, known as mass transport models (MTMs). In these models, mass—either discrete or continuous—is transferred between neighboring sites according to specified stochastic {rules~\cite{Spitzer,Stinchcombe,Stinchcombe,MTM}}. The transfer rates can depend on the occupation numbers of the sites involved, enabling a variety of local interaction effects to be incorporated into the system dynamics. MTMs have been applied to study transport, aggregation, and redistribution processes in both physical and biological {systems~\cite{MTMAPP,Sachdeva2013}. }

A well-studied example of driven lattice systems is the asymmetric simple exclusion process  {(ASEP)~\cite{Spitzer,Stinchcombe, ASEP}}, in which particles hop to neighboring sites under a hard-core exclusion constraint, allowing at most one particle per site. The ASEP has been used extensively as a minimal model to investigate driven transport under non-equilibrium conditions. Another prominent model is the zero-range process  {(ZRP)~\cite{Spitzer,Stinchcombe,ZRP}}, which permits arbitrary occupation numbers at each site and features single-particle hopping rates that depend solely on the occupation of the departure site. The ZRP is analytically tractable and admits a steady-state distribution of product form. These features of {the} ZRP have made it a widely used model,  {including for} analyzing condensation transitions in a variety of {systems~\cite{ZRP2}.}

{ In} the current work, we study an MTM defined on a lattice, and focus on the interplay between interparticle repulsion and a biasing field which preferentially drives particles in one direction. Our model bridges features of the ASEP and ZRP. {An interesting point is that while the ZRP arises straightforwardly in the limit of vanishing repulsion, the ASEP arises as an emergent property of the state at large values of the repulsion.} 

{The model is defined on a lattice} with discrete particles where multiple occupancy is permitted, but with an additional on-site repulsion energy \(E(n_{i})\) given by
\begin{align}
    E(n_{i})=U\frac{n_{i}(n_{i}-1)}{2},\quad U\geq0,\label{onsite}
\end{align} 
where $n_i$ is the occupancy at site $i$ and $U$ is a repulsive on-site interaction. Further, particles have asymmetric rates for hopping in different directions, with the ratio of forward to backward  rates being $\frac{(1+g)}{(1-g)}$. The model differs from the ZRP by including arrival-site dependence, and from the ASEP by allowing multiple occupancy. {The form of the on-site repulsion term (Eq.~\eqref{onsite}) is the same as in the quantum Bose-Hubbard model, which describes interacting bosons on a lattice~\cite{BHM,DipoleCond,Russ2025,Yan2017}. Our study may of relevance to this system in the classical (high temperature) regime. {In view of the above mentioned similarity}, we use the nomenclature `Bose-Hubbard' to refer to the interactions in Eq.~\eqref{onsite}.}

\begin{figure}[h!]
\centering
\centering
\begin{tikzpicture}[scale=6]
  \fill[orange!40, opacity=0.5] (0,0) rectangle (1,0.05);

  \fill[blue!20, opacity=0.3] (0,0.05) rectangle (1,0.2);

  \fill[red!20, opacity=0.3] (0,0.2) rectangle (1,0.5);

  \fill[blue!20, opacity=0.3] (0,0.5) rectangle (1,0.8);

  \fill[violet!30, opacity=0.3] (0,0.8) rectangle (1,1);

  \fill[green!30, opacity=0.3] (0,0.05) rectangle (0.2,0.7);

  \draw[->, thick] (0,0) -- (1.1,0);
  \draw[->, thick] (0,0) -- (0,1.1);

  \node at (0.55,-0.05) {{Bias} $\displaystyle g$};
  \node[rotate=90] at (-0.07,0.55) {{Repulsion} $\displaystyle{\frac{\beta U}{1+\beta U}}$};

  \draw[thick] (0,0) -- (1,0) -- (1,1) -- (0,1) -- cycle;

  \node[below left] at (0,0) {$\displaystyle{(0,0)}$};
  \draw (1,0.01) -- (1,-0.01) node[below] {$\displaystyle{1}$};
  \draw (0.01,1) -- (-0.01,1) node[left] {$\displaystyle{1}$};

  \node at (0.5, 0.125) {\footnotesize \textbf{Low Correlation}};
  \node at (0.5, 0.35) {\footnotesize \textbf{High Correlation}};
  \node at (0.5, 0.6) {\footnotesize \textbf{Low Correlation}};
  \node at (0.5, 0.88) {\footnotesize \parbox{2.5cm}{\centering \textbf{\textit{Emergent ASEP}}}};

  \node[rotate=90] at (0.1, 0.35) {\footnotesize \textbf{Linear Response}};

  \node at (0.5, 0.025) {\footnotesize \textbf{\textit{ZRP Limit}}};
\end{tikzpicture}

     \caption{Qualitative phase diagram showing different regimes in a one-dimensional periodic lattice.{ On the horizontal axis, the system reduces to a ZRP, while on the vertical axis, it has an equilibrium product measure steady state. }The diagram illustrates regions of low and high correlation, the {emergent} ASEP regime, and the domain where linear response theory is valid. Boundaries between these regimes are smooth crossovers rather than sharp transitions. The {emergent} ASEP limit is formally reached when $\beta U \to \infty$, but its characteristic behavior appears already for $\beta U \gg 1$.}
\label{fig:correlation-diagram}

\end{figure}

To investigate the steady-state transport properties and density structures emerging from the interplay of interactions and biased stochastic dynamics, we study a one-dimensional lattice model for both periodic and {free} boundary conditions. {Free boundary conditions are closer to experimental realizations in optical lattices~\cite{BHM,Yan2017,DipoleCond,Russ2025}}. Periodic and free boundary conditions lead to markedly different steady-state behaviors. 

With periodic boundary {conditions}, we analyze steady-state observables such as particle current and spatial correlations, focusing on their dependence on particle density, interaction strength $\beta U$, and biasing parameter. {Interestingly} in the limit of large $\beta U$, the steady-state {consists of stacks of particles over which ride single particles and holes, whose dynamics is described by the ASEP. The pattern is repeated as the density is increased and as a consequence the} current exhibits a periodic dependence on density. Furthermore, inter-site correlations show non-monotonic behavior with respect to $\beta U$: they vanish both in the limit of infinitesimal $\beta U$ and infinite $\beta U$, {with a maximum in between}. The different regimes arising in a one-dimensional system with periodic boundary conditions (1D PBC) are summarized in the phase diagram shown in Fig.~\ref{fig:correlation-diagram}. 

With {free} boundary conditions---relevant to experimental realizations of BHMs in tilted lattices---the competition between the interaction-induced repulsion and the external bias gives rise to an interesting behavior. In particular, at low temperatures, the system develops multiple plateau-like density profiles that are sensitive to both the interaction strength and the applied bias. As the bias increases, the density profile undergoes a transition characterized by the emergence of a macroscopic region of empty sites, followed by a steep rise in density. 

In summary, our model enables the exploration of non-equilibrium transport in systems with local interactions, biased hopping, and varying boundary conditions. By combining features of the ASEP, ZRP, and the Bose–Hubbard model (BHM), {we uncover several varied and interesting regimes of behaviour in the steady state. }

%% file: sections/section02.tex
\section{Model} \label{sec:model}

We consider a system of identical particles on a lattice $\Lambda$ consisting of $L$ sites, indexed by $i = 1, 2, \dots, L$. The number of particles at site $i$ is denoted by $n_i \in \mathbb{Z}_{\geq 0}$, and a configuration of the system is described by the vector $\mathbf{n} \equiv (n_1, n_2, \dots, n_L)$. We also fix the total density $\rho$:
\begin{align}
    \sum_{i=1}^{L}n_{i}=N,\quad\frac{N}{L}=\rho. 
\end{align}The dynamics of the system is governed by stochastic particle hops between nearest-neighbor sites, influenced by a directional bias and an energy-based transition probability.
\renewcommand{\figurename}{FIG.}

\begin{figure}[h!]
\centering
\resizebox{\columnwidth}{!}{%
\begin{tikzpicture}[scale=1.0, every node/.style={scale=0.8}, >=Latex]  

\def\xspace{1.2}

\foreach \x in {1,2,3,4,5,6} {
    \pgfmathsetmacro{\xcoord}{\x*\xspace}
    \draw (\xcoord,0) -- (\xcoord+0.5,0);
    \node[below] at (\xcoord+0.25,0) {\textbf{\x}};
}

\node[below left, font=\bfseries] at (0.8*\xspace,0) {Site:};

\foreach \y in {1,2,3} {
    \draw (2*\xspace+0.25,0.3*\y) circle (0.12);
}
\fill (2*\xspace+0.25,0.3*4) circle (0.12); 

\foreach \y in {1,2,3} {
    \draw (3*\xspace+0.25,0.3*\y) circle (0.12);
}

\fill (4*\xspace+0.25,0.3) circle (0.12);

\draw (5*\xspace+0.25,0.3*1) circle (0.12);
\fill (5*\xspace+0.25,0.3*2) circle (0.12);

\foreach \y in {1,2} {
    \draw (6*\xspace+0.25,0.3*\y) circle (0.12);
}

\draw[->, thick, >=Latex] 
    (2*\xspace+0.25,1.5) 
    to[out=120, in=60] 
    node[midway, yshift=20pt, scale=0.9] {\textit{\textbf{$W_2(4,0)$}}}
    (1*\xspace+0.25,0.3);

\draw[->, thick, >=Latex] 
    (4*\xspace+0.25,0.6) 
    to[out=120, in=60] 
    node[midway, yshift=12pt, scale=0.9] {\textit{\textbf{$W_2(1,3)$}}}
    (3*\xspace+0.25,1.2);

\draw[->, thick, >=Latex] 
    (5*\xspace+0.25,0.9) 
    to[out=45, in=135] 
    node[midway, yshift=10pt, scale=0.9] {\textit{\textbf{$W_1(2,2)$}}}
    (6*\xspace+0.25,0.9);

\end{tikzpicture}

}
\caption{
Illustration of stochastic particle dynamics in a one-dimensional lattice with six sites. Each circle represents a particle at a site, and arrows denote possible hopping transitions of a single particle between nearest-neighbor sites. The particle which attempts a move is shaded for reference. The transition rates $W_1(n_i, n_j)$ and $W_2(n_i, n_j)$ depend on the occupation numbers $n_i$ and $n_j$ of the departure and arrival sites, respectively. For instance, $W_2(4,0)$ represents the hopping rate of a particle moving from site 2 (with 4 particles) to site 1 (empty).    
}
\label{fig:stochastic_hopping}
\end{figure}
We outline the dynamics on a 1D lattice. In each microscopic time step, we select a bond at random on the lattice with uniform probability $\frac{1}{L}$. $L$ such microscopic steps {constitute one} Monte-Carlo step. Next, the direction of attempted motion across that bond is chosen probabilistically. A particle is selected to move from site \(i\) to site \(i+1\) with probability \((1+g)/2\), and from site \(i+1\) to site \(i\) with probability \((1-g)/2\). The parameter \(g \in [0,1]\) controls the degree of directional bias: \(g = 0\) corresponds to symmetric (unbiased) dynamics, while \(g = 1\) represents fully asymmetric dynamics favoring forward motion.

A particle—if present—attempts to hop from the departure site to its neighboring site. If {the} system is in configuration $
    \mathcal{C} \equiv \mathbf{n},
$
and a hop is attempted across the bond $(i, i+1)$ from site $i$ to site $i+1$. A successful move results in a configuration
$
    \mathcal{C}^{{\prime}} \equiv \mathbf{n}_{i,i+1}^{-,+},
$
where $\mathbf{n}_{i,j}^{\pm,\mp}$ denotes the configuration obtained by changing $n_i$, $n_{j}$ by $\pm1$, and $\mp1$, respectively. Assuming an on-site energy given by Eq.~\eqref{onsite}, the change in energy is given by
\begin{align}
\Delta E(\mathcal{C} \to \mathcal{C}^{{\prime}}) = U(n_{i+1} - n_i + 1).
\end{align}
This move is accepted with a Metropolis-type probability:
\begin{align}
p = \min\left(1, \exp\left(-\beta \Delta E(\mathcal{C} \to \mathcal{C}^{\prime})\right)\right), \label{p}
\end{align}
where $\beta \geq 0$ denotes an inverse temperature parameter that modulates the system’s sensitivity to energy differences.

Analogously, for a single particle hop across the bond $(i, i+1)$ from site $i+1$ to $i$, the system transitions from configuration $\mathcal{C}$ to $\mathcal{C}^{\prime\prime} \equiv \mathbf{n}_{i,i+1}^{+,-}$, if the move is successful. The energy change associated with this move is
{
\begin{align}
\Delta E(\mathcal{C} \to \mathcal{C}^{\prime\prime}) = U(n_i - n_{i+1} + 1),
\end{align}}and a corresponding Metropolis acceptance probability is constructed similarly to Eq.~\eqref{p} with the result
\begin{align}
q = \min\left(1, \exp\left(-\beta \Delta E(\mathcal{C} \to \mathcal{C}^{\prime\prime})\right)\right). \label{q}
\end{align}

This stochastic dynamics can equivalently be described within a master equation framework. Let $P(\mathbf{n}; t)$ denote the probability of the system being in configuration $\mathbf{n}$ at time $t$. On a one-dimensional periodic lattice, the hopping rates associated with the dynamical rules above are given by
\begin{align}
W_1(\mathbf{n} \to \mathbf{n}_{i,i+1}^{-,+}) &= p\,\frac{1+g}{2}\, \theta(n_i), \notag \\
W_2(\mathbf{n} \to \mathbf{n}_{i,i+1}^{+,-}) &= q\,\frac{1-g}{2}\, \theta(n_{i+1}).\label{hop} 
\end{align}
In Eq.~\eqref{hop}, \( W_1 \) represents the net hopping rate of a particle from site \( i \) to site \( i+1 \), while \( W_2 \) denotes the net hopping rate from site \( i+1 \) to site \( i \). The Heaviside theta functions ensure that hops occur only when the departure site has at least one particle. Also, since the transition rates depend only on the occupancy of the departure and arrival sites, it is convenient to write them as $W_{1}(m, n)$ and $W_{2}(m, n)$, where $m$ is the occupation at the departure site and $n$ at the arrival site. Single particle hopping moves at various sites are illustrated in Fig.~\ref{fig:stochastic_hopping}. 

We focus on the non-equilibrium steady-state (NESS) properties of the system. These steady states arise from the structure of the $\mathbb{W}$ matrix in the master equation~\cite{van}, and are defined by time-independent configuration probabilities $P(\mathbf{n}; t)$, meaning that $\frac{d}{dt}P(\mathbf{n}; t) = 0$ for all configurations $\mathbf{n}$. The corresponding master equation is

{\small
\begin{align}
\frac{d}{dt} P(\mathbf{n}; t) 
&= \sum_{i} \Big[
    W_1(n_{i-1}+1, n_i-1)\, 
    P(\mathbf{n}_{i-1,i}^{+,-}; t) \nonumber\\
&\quad 
  + W_2(n_{i+1}+1,n_i-1)\, 
    P(\mathbf{n}_{i,i+1}^{-,+}; t) \nonumber\\
&\quad 
  - P(\mathbf{n}; t) 
    \left( W_1(n_i, n_{i+1})+W_2(n_i,n_{i-1})  \right)
\Big] \theta(n_i).
\label{me}
\end{align}
}
    
    {The steady states are sensitive to boundary conditions. In Sec.~\ref{sec:PBC} below, we consider periodic boundary conditions, which ensure the translational invariance of density. In Sec.~\ref{sec:OBC}, we study the system with free boundary conditions, in which case, there is no current, and the steady state is an equilibrium state. In {a certain parameter range, we find that the density profile shows pronounced steps along the lattice}} 
    
  {In a companion paper~\cite{RC} on Bose-Hubbard systems in the presence of bias and disorder, we study the effect of a different sort of microscopic dynamics which also incorporates local detailed balance and yields an analytically characterized steady state. With periodic boundary conditions, the steady state is quite different from the present work, but with free boundary conditions, it is an identical equilibrium state.} 

%% file: sections/section03.tex
\section{The Periodic Ring} \label{sec:PBC}
In this section, we consider the time evolution of the configuration probabilities $P(\mathbf{n}; t)$ on a one-dimensional lattice with periodic boundary conditions, $n_{i+L} = n_i$. 

The steady state of Eq.~\eqref{me} can be obtained exactly in the following limiting cases: $\beta U \to 0$, and $\beta U \to \infty$ for arbitrary values of the asymmetry parameter $g$, and in the case $g\to0$ for arbitrary values of $\beta U$. For intermediate interaction strengths, we employ Monte Carlo simulations using the Metropolis algorithm, as described in Sec.~\ref{sec:model}.

In the steady state, we focus on two key observables: the two-point connected correlation function $G(r)$ and the steady-state current $j$, defined respectively as
\begin{align}
    G(r) = \langle n_i n_{i+r} \rangle - \langle n_i \rangle \langle n_{i+r} \rangle = \langle n_i n_{i+r} \rangle - \rho^2,
    \label{G(r)}
\end{align}
\begin{align}
    j = \left\langle W_1(n_i, n_{i+1}) - W_2(n_{i+1}, n_i) \right\rangle.
    \label{j}
\end{align}
In Eq.~\eqref{G(r)} we have used translational invariance to write the last equality. Also, in Eq.~\eqref{j}, there is no {prefactor} before the weights, since only one particle moves in an elementary hop.
\subsection{Limiting Cases}
\subsubsection{Zero-range-like limit ($\beta U \to 0$)}  In this limit, all configurations with a fixed total particle number $N$ are equally probable~\cite{ZRP}, and the model reduces to a ZRP. The steady-state distribution is given by
\begin{align}
    P(\mathbf{n}) = \frac{1}{\binom{N + L - 1}{L - 1}} \, \delta\left( \sum_{i=1}^{L} n_i - N \right),
\end{align}
where the Kronecker delta ensures particle number conservation. In the thermodynamic limit ($L, N \to \infty$ with fixed density $\rho = N/L$), the steady-state current, which is determined by the average number of occupied sites—since particles can only hop from occupied sites—is given by
\begin{align}
    j^0 = g \frac{\rho}{1 + \rho}. \label{j0}
\end{align}
This current saturates at $j^0 \leq g$ as $\rho \to \infty$. For any finite interaction strength $\beta U$, the current is bounded above by this non-interacting result (Fig.~\ref{fig:current_vs_g}), as repulsive interactions further inhibit particle transitions:
\begin{align}
    j(\beta U, \rho, g) \leq j^0(\rho, g).
\end{align}

The connected correlation function $G(r)$ vanishes in the thermodynamic limit for all $r \neq 0$. 
\subsubsection{Strongly interacting limit (\texorpdfstring{$\beta U \to \infty$}{beta U goes to infinity})}
  In the strongly interacting regime, large interaction energies effectively suppress transitions that would lead to highly occupied sites. As a result, the system dynamically projects onto a restricted configuration space with the majority of particles {being immobile}, and the left over particles (at most one per site) exhibiting quasi-hard-core behavior, thereby mapping into an effective ASEP.

For a fixed density $\rho$, the energy-minimizing configuration corresponds to a uniform background where each site is occupied by $\lfloor \rho \rfloor$ particles (the integer part of $\rho$), with the remaining particles (i.e., the excess) distributed such that no site receives more than one additional particle. Thus, the occupation number at each site $i$ can be written as
\begin{align}
    n_i = \lfloor \rho \rfloor + \tilde{m}_i, \quad \tilde{m}_i \in \{0, 1\},
\end{align}
where $\tilde{m}_i = 1$ indicates the presence of an excess (or "quasi-hard-core") particle above a uniform background {of stacked particles.}

We denote by $\tilde{m}$ and $\tilde{n}$ the values of this excess variable at the departure and arrival sites, respectively, during a hopping event. In this limit, a particle can only hop if an excess particle ($\tilde{m} = 1$) moves to a site that does not already host one ($\tilde{n} = 0$). Hopping processes that violate this constraint are energetically forbidden.
 In the strong-interaction limit, the energetic cost of such a process becomes
\begin{align}
    \lim_{\beta U \to \infty} \beta \Delta E \to
    \begin{cases}
        0 & \text{if } \tilde{m} = 1,\ \tilde{n} = 0, \\
        \infty & \text{otherwise}.
    \end{cases}
\end{align}
The effective transition rates in this regime are:
\begin{align}
    W_1(\mathbf{n} \to \mathbf{n}_{i,i+1}^{-,+}) &= \frac{1 + g}{2} \tilde{m}_i (1 - \tilde{m}_{i+1}), \nonumber \\
    W_2(\mathbf{n} \to \mathbf{n}_{i,i+1}^{+,-}) &= \frac{1 - g}{2} \tilde{m}_{i+1} (1 - \tilde{m}_i).
\end{align}

The steady-state distribution over the reduced configuration space $\mathcal{C} = (\tilde{m}_1, \tilde{m}_2, \dots, \tilde{m}_L)$ is uniform:
\begin{align}
    P(\tilde{m}_1, \dots, \tilde{m}_L) = \frac{1}{\binom{L}{M}} \, \delta\left( \sum_{i=1}^{L} \tilde{m}_i - M \right),
\end{align}
where the number of quasi-particles is
\begin{align}
    M = \sum_{i=1}^{L} n_i - L \lfloor \rho \rfloor.
\end{align}
Since there is a uniform stacking of particle till the greatest integer function of $\rho$, and it is the excess particles which perform the dynamics, the current $j^{\infty}(\rho,g)$ is periodic in $\rho$ {with period unity} and is given by
\begin{align}
    j^\infty = g\, \tilde{\rho}(1 - \tilde{\rho}), \quad \text{where } \tilde{\rho} = \frac{M}{L}.\label{jin}
\end{align}
The oscillations in $j^\infty$ as a function of $\rho$ are also present 
for large but finite values of $\beta U$, but the amplitude 
decreases as the value of $\beta U$ reduces (Fig.~\ref{fig:jp1}).

This current $j^\infty$ serves as a lower bound(Fig.~\ref{fig:current_vs_g}) for $j(\beta U,\rho,g)$ for a fixed $g,\rho$:
\begin{align}
    j^\infty(\rho, g) \leq j(\beta U,  \rho,g) \leq j^0(\rho, g)\leq g.\label{bound}
\end{align}
In this limit, the drift velocity $ v_d = \frac{j}{\rho} $ exhibits an oscillatory dependence on the density and decreases in magnitude with increasing density. This behavior arises {because} at higher densities, the probability of selecting a specific particle at random decreases.

 The connected correlation function $G(r)$ vanishes for $r \neq 0$ in the thermodynamic limit. The on-site variance $G(0)$ is unit-periodic in $\rho$ and independent of $g$:
\begin{align}
    G(0) = \tilde{\rho}(1 - \tilde{\rho}).\label{gin}
\end{align}
Thus, the system maps onto an effective ASEP characterized by oscillatory variance and current, {and a} damped oscillatory drift velocity.
\subsection{Numerical Results}
For intermediate values of $\beta U$ and $g$, we numerically analyze $j$, $G(0)$, and $G(1)$ defined in Eqs.~\eqref{j} and~\eqref{G(r)} respectively. The algorithm has been outlined in Sec.~\ref{sec:model}. 
\subsubsection{Current} {The current $j$ is} the average net hopping rate across any bond in the steady state. In general, $j$ is a function of interaction strength $\beta U$, the particle density $\rho$, and the bias parameter $g$.

\renewcommand{\figurename}{FIG.}
\begin{figure}[h!]
    \centering
    \includegraphics[width=0.48\textwidth]{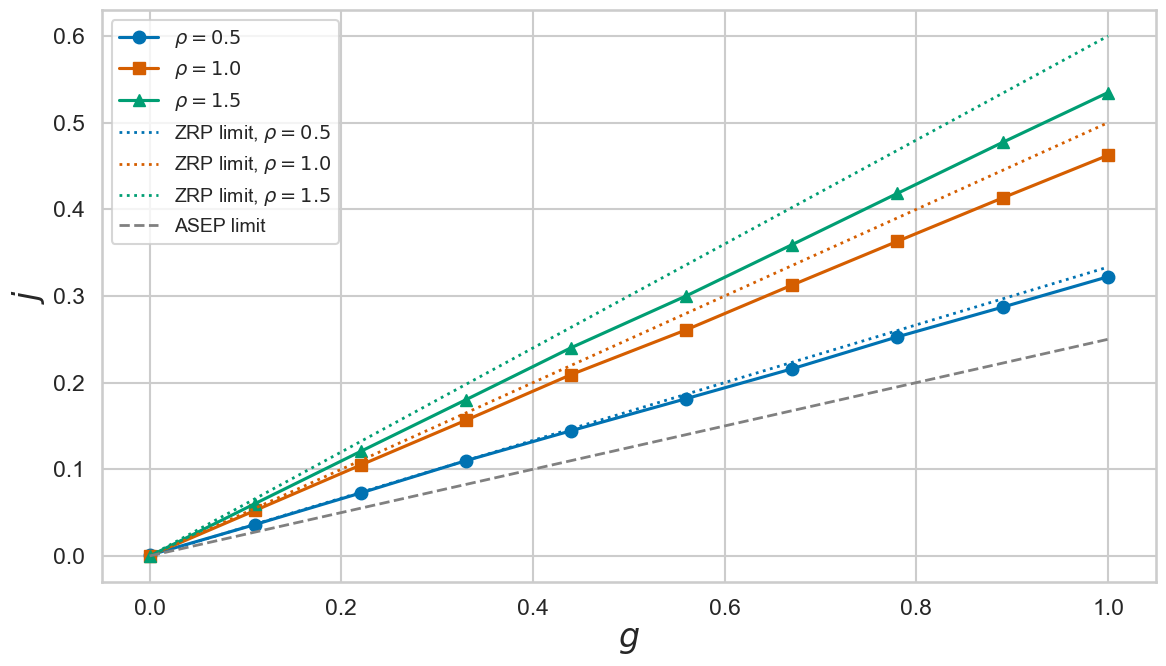}
    \caption{Steady-state current $j$ as a function of the biasing parameter $g$ for different values of $\rho$ at $\beta U = 0.5$. Other simulation parameters: System size $L=150$, $10^5$ Monte Carlo steps, $5L^2$ relaxation time. {The dashed black line correspond to the emergent ASEP limit for fractional part of $\rho$ fixed at 0.5. The ZRP and the ASEP limits serve as upper and lower bounds of the current respectively.}}
    \label{fig:current_vs_g}
\end{figure}

\begin{figure}[h!]
    \centering
    \includegraphics[width=0.48\textwidth]{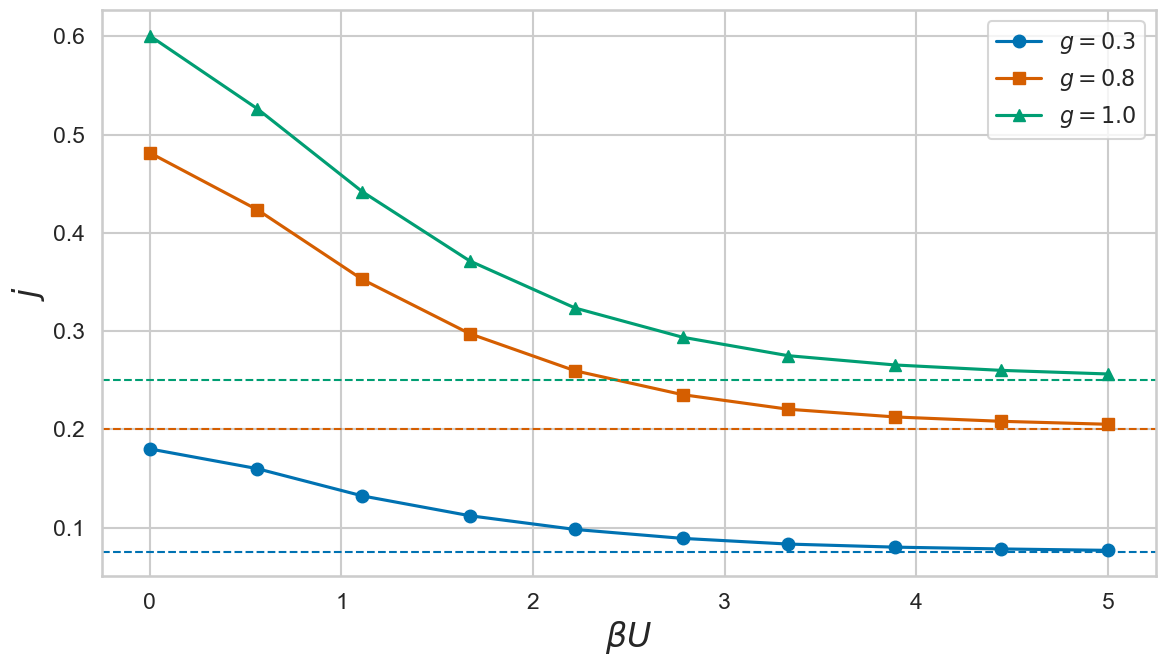}
    \caption{Steady-state current $j$ as a function of the interaction strength $\beta U$ for different values of $g$ at fixed $\rho = 1.5$. Other simulation parameters, as for Fig.~\ref{fig:current_vs_g}. Dashed lines indicate the limiting value of $j$ at large $\beta U$.}
    \label{fig:current_vs_u}
\end{figure}

Figure~\ref{fig:current_vs_g} illustrates how the current $j$  varies with the biasing parameter $g$. As expected, increasing $g$ leads to a higher net current, as the backward hopping rate becomes increasingly suppressed. At fixed interaction strength $\beta U = 0.5$, a higher density generally leads to an increase in current, since more particles are available to contribute to transport. However, this trend does not always hold true: at larger interaction strengths, increased crowding leads to enhanced particle repulsion, which can make the current non-monotonic in density, as we will discuss below in the context of Fig.~\ref{fig:current_vs_u}.

\renewcommand{\figurename}{FIG.}
\begin{figure}[h!]
    \centering
    \begin{subfigure}[b]{0.48\textwidth}
        \includegraphics[width=\textwidth]{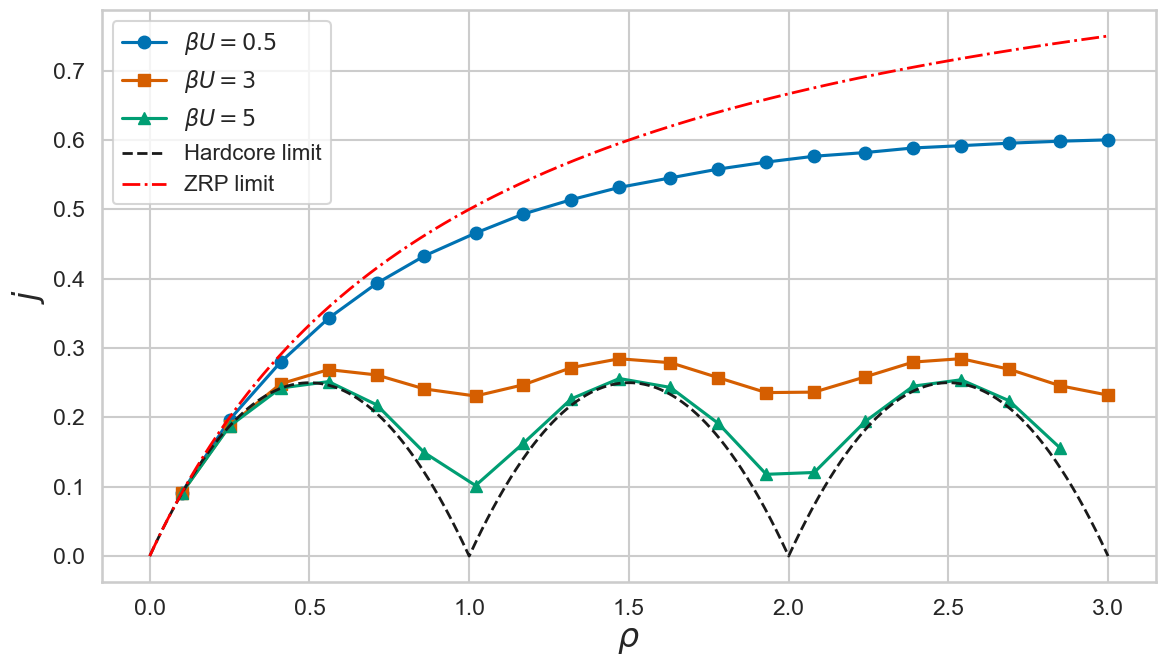}
        \caption{Current $j$ versus $\rho$ for different values of $\beta U$, at fixed $g=1$.}
        \label{fig:jp1}
    \end{subfigure}
    \hfill
    \begin{subfigure}[b]{0.48\textwidth}
        \includegraphics[width=\textwidth]{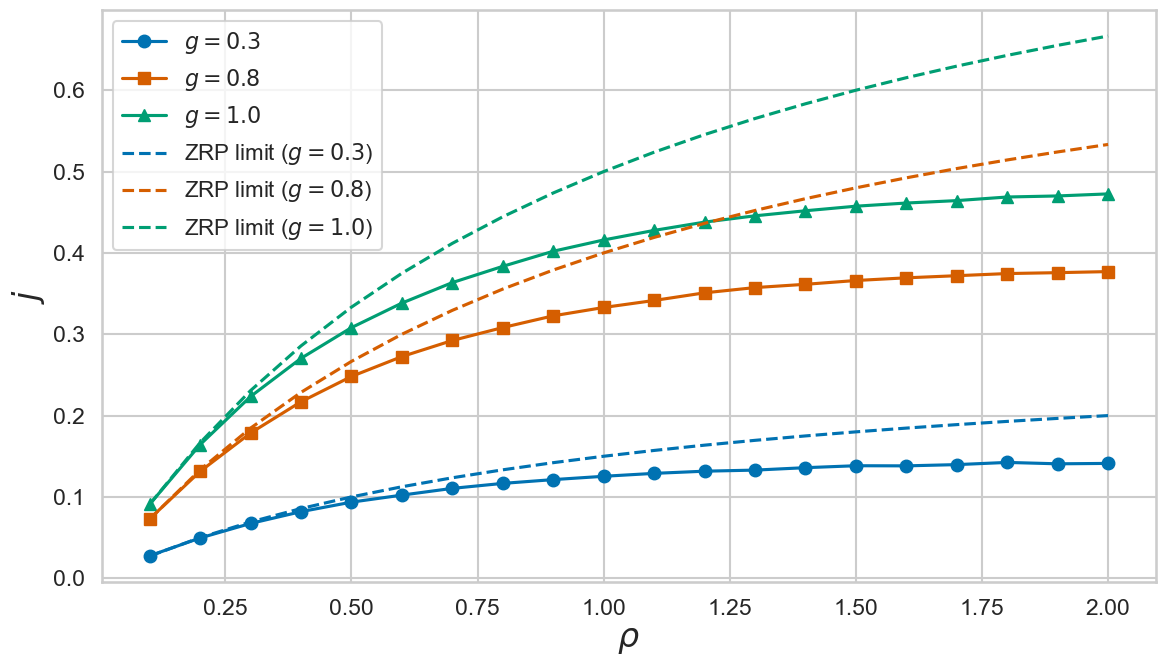}
        \caption{Current $j$ versus $\rho$ for different values of $g$, at fixed $\beta U = 1$.}
        \label{fig:jp2}
    \end{subfigure}
    \caption{Steady-state current $j$ as a function of density $\rho$ for varying biasing parameter $g$ and interaction strength $\beta U$. Other simulation parameters, as for Fig.~\ref{fig:current_vs_g}. Dashed lines indicate the limiting case(s).}
    \label{fig:current_vs_p}
\end{figure}

Fig.~\ref{fig:jp1} shows that for large $\beta U$, the current exhibits an oscillatory dependence on $\rho$, as predicted by the analytic expressions for $j$ in the strong interaction regime (Eq.~\eqref{jin}). The transition from smooth saturation to oscillatory behavior occurs as $\beta U$ increases. In Fig.~\ref{fig:jp2} for fixed $g = 1$, we observe that at low to intermediate $\beta U$, the current initially increases with density, but eventually saturates due to interaction-induced hindrance of particle mobility. 

\renewcommand{\figurename}{FIG.}

 Figure~\ref{fig:current_vs_u} reveals that for fixed $g$, the current eventually saturates or even decreases as $\beta U$ becomes large. Notably, in the strong interaction limit, only the fractional part of the density $\rho$ contributes to the current. Densities differing by an integer but sharing the same fractional component converge to the same current value, whereas the current decays to zero for exact integer densities. This is consistent with the effective exclusion behavior induced by strong on-site repulsions. 

In summary, our numerical results demonstrate how the steady-state current $j$ is shaped by the interplay between interaction strength $\beta U$, particle density $\rho$, and directional bias $g$. Increasing $g$ monotonically enhances the current by suppressing backward hopping. In contrast, increasing $\beta U$ introduces an interesting behavior: at low and intermediate interaction strengths, current increases with density but eventually saturates; at high interaction strengths, it exhibits oscillations tied to the fractional part of the density.
\subsubsection{Correlation functions}
{We study the correlation function defined in Eq.~\eqref{G(r)}, for $r=0,1.$}
\begin{center}
    \textit{{(a) On-site variance $G(0)$}}
\end{center}
The on-site variance $G(0)$ is defined as
\begin{align}
    G(0) = \langle n_i^2 \rangle - \rho^2,
\end{align}
where $\langle n_i^2 \rangle$ is the local second moment of the particle number.
\renewcommand{\figurename}{FIG.}
\begin{figure}[h!]
    \centering
    \includegraphics[width=0.48\textwidth]{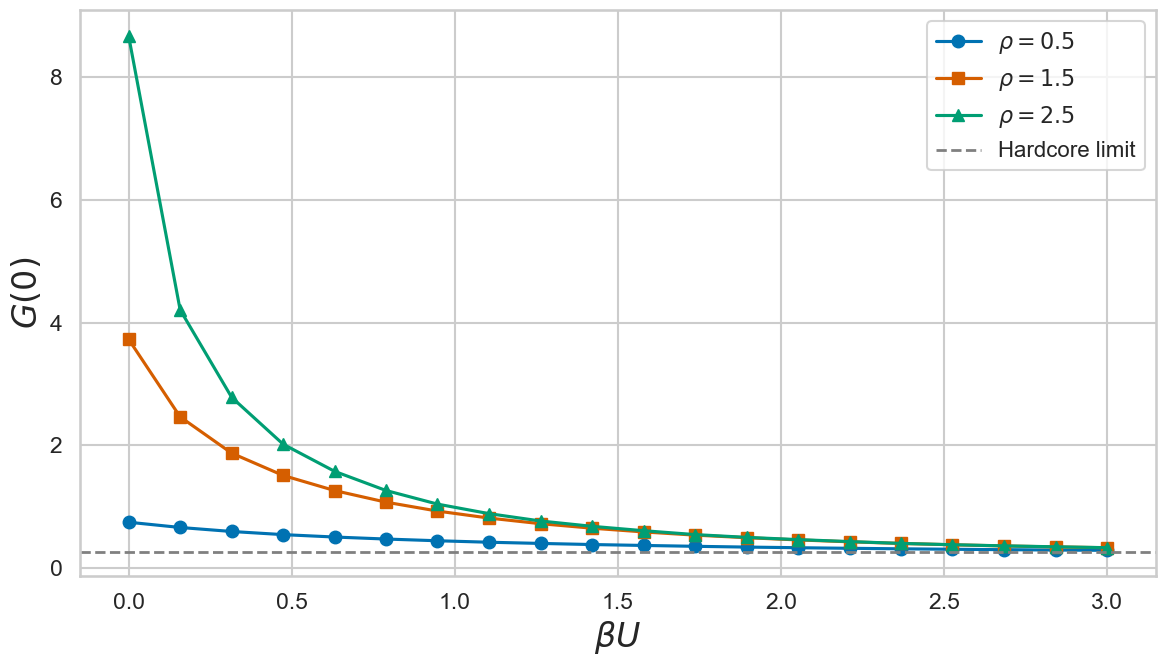}
    \caption{On-site variance $G(0)$ versus $\beta U$ for different values of $\rho$ at fixed $g = 1$. Other simulation parameters, as for Fig.~\ref{fig:current_vs_g}. Dashed lines denote the hard-core limit as given by Eq.~\eqref{gin}.}
    \label{fig:G0_combined0}
\end{figure}

Figure~\ref{fig:G0_combined0} shows the dependence of the on-site variance $G(0)$ on the interaction strength $\beta U$. {We observe that} $G(0)$ for different densities $\rho$ with identical fractional parts exhibit convergence to the same asymptotic value at large $\beta U$, consistent with theoretical predictions for the hard-core limit. The observed decrease in $G(0)$ with increasing $\beta U$ can be attributed to the suppression of fluctuations as the system becomes more ordered due to increased repulsion, leading to particle stacking and variance saturation.

\renewcommand{\figurename}{FIG.}
\begin{figure}[h!]
    \centering
    \includegraphics[width=0.48\textwidth]{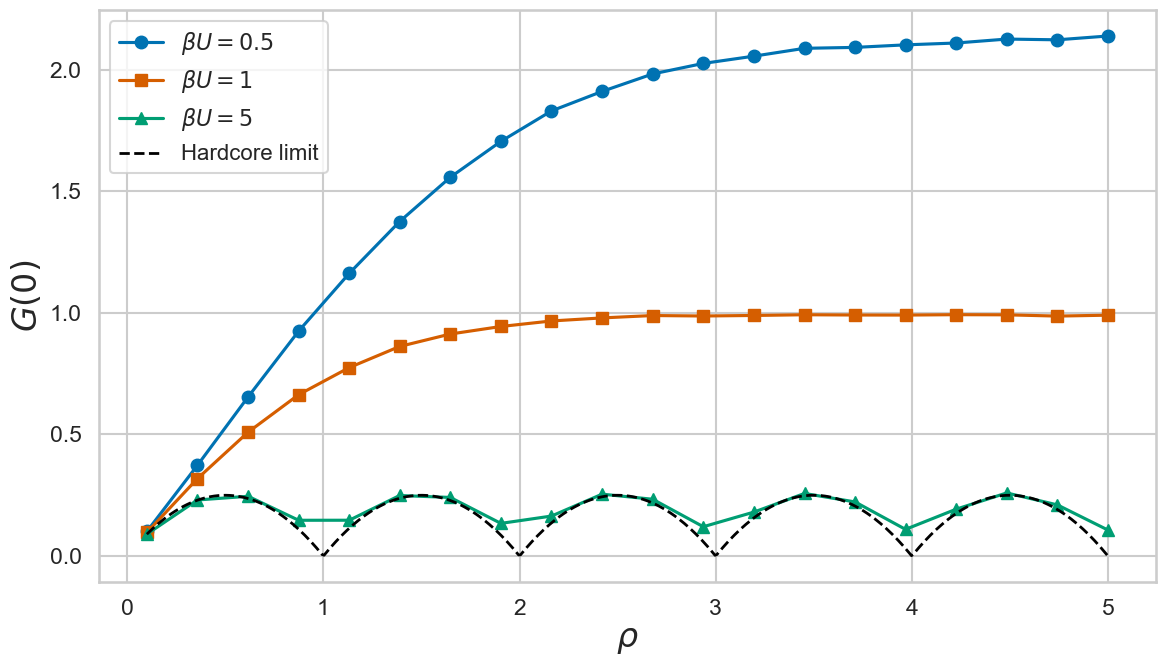}
    \caption{On-site variance $G(0)$ versus $\rho$ for different values of $\beta U$ at fixed $g = 1$. Other simulation parameters, as for Fig.~\ref{fig:current_vs_g}. Dashed lines denote the hard-core limit as given by Eq.~\eqref{gin}.}
    \label{fig:G0_combined}
\end{figure}

Fig.~\ref{fig:G0_combined} shows that the variance is monotonic in density for small $\beta U$, however for large $\beta U$, we again recover the oscillatory nature. This is because at small values of {$\beta U$,} the system can fluctuate more due to low interactions strength, which would be more prominent for higher densities, with eventual saturation due to higher magnitudes of repulsive energy at higher densities. For $\beta U\gg1$, the ASEP nature is recovered and the variance follows Eq.~\eqref{gin}.
\begin{center}
 \textit{   {(b)Nearest neighbor correlations}}
\end{center} {The correlation function for adjacent sites is}
\begin{align}
    G(1) = \langle n_in_{i+1} \rangle - \rho^2.
\end{align}

\renewcommand{\figurename}{FIG.}
\begin{figure}[h!]
    \centering
    \begin{subfigure}[b]{0.48\textwidth}
        \includegraphics[width=\textwidth]{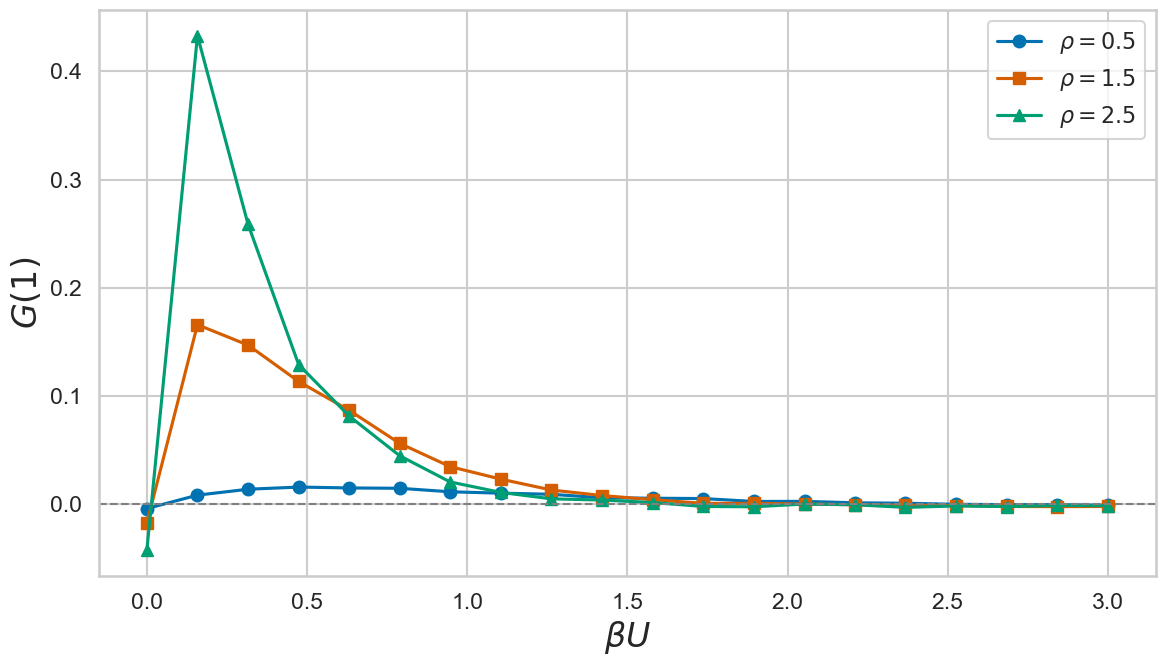}
        \caption{Nearest neighbor correlation $G(1)$ versus $\beta U$ for different values of $\rho$ at fixed $g=1$.}
        \label{fig:G0U1}
    \end{subfigure}
    \hfill
    \begin{subfigure}[b]{0.48\textwidth}
        \includegraphics[width=\textwidth]{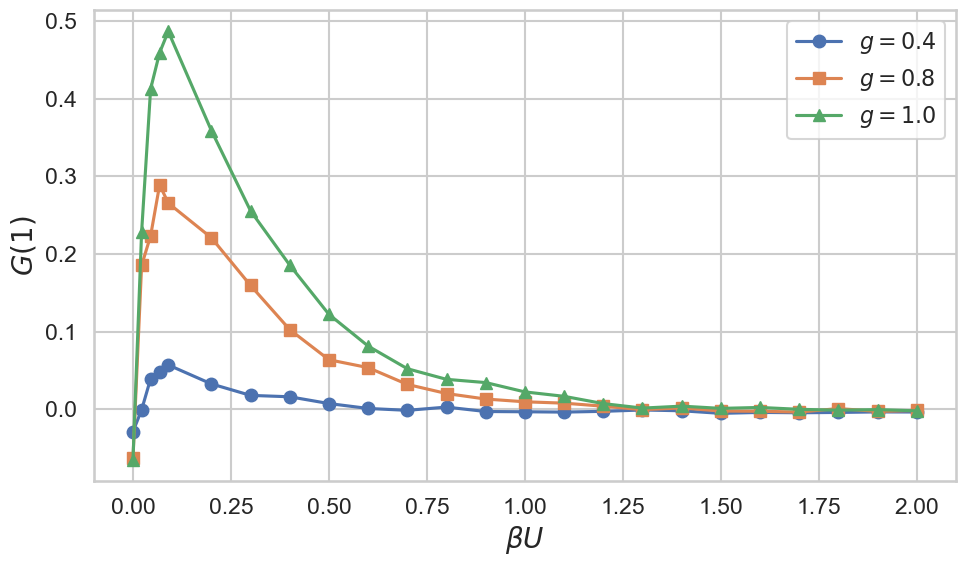}
        \caption{Nearest neighbor correlation $G(1)$ versus $\beta U$ for different values of $g$ at fixed $\rho=2.5$.}
        \label{fig:G0U2}
    \end{subfigure}
    \caption{Nearest neighbor correlation $G(1)$ as function of the interaction strength $\beta U$, for varying density $\rho$ and biasing parameter $g$. Other simulation parameter, as for Fig.~\ref{fig:current_vs_g}}
    \label{fig:G1_combined}
\end{figure}

Figure~\ref{fig:G1_combined} illustrates the non-monotonic dependence of \( G(1) \) on the interaction strength \(\beta U\). This can be understood as follows: in both limiting cases—non-interacting (\(\beta U \to 0\)) and strongly interacting (\(\beta U \to \infty\))—the correlation function vanishes. Therefore, if any correlation exists, it must exhibit a maximum at an intermediate value of the interaction strength.

In general, the hopping process considered here involves rates that depend on both the departure and arrival sites. As shown in~\cite{MP}, product measure steady states are only possible if the hopping rates satisfy specific constraints. However, the rates defined in Eq.~\eqref{hop} do not satisfy these conditions, implying the presence of spatial correlations.

Furthermore, \( G(1) \) increases monotonically with the biasing parameter \( g \), interpolating between an equilibrium state at \( g = 0 \) and a fully biased non-equilibrium steady state (NESS) at \( g=1 \).

This non-monotonic behavior is not restricted to \( G(1) \) alone, but is also observed for correlations at larger separations, \( G(r) \) with \( r > 1 \). However, these correlations are typically smaller in magnitude compared to \( G(1) \). Interestingly, the overall shape of \( G(r) \) as a function of \(\beta U\), as well as the location of its maximum, remains nearly unchanged across different \( r \), provided other parameters are held constant.

 {The presence of nearest neighbor correlations shows that the particles begin to cluster, that is, a particle at one site increases the likelihood of finding a particle at a neighboring site.}
\subsubsection{Linear Response Theory}
In the limit $g \to 0$, the Metropolis Monte Carlo dynamics ensures that the system relaxes to the Boltzmann-Gibbs distribution. In this equilibrium limit, the probability of a configuration $\mathbf{n}$ is given by
\begin{align}
    P(\mathbf{n}) = \frac{1}{\mathcal{Z}} \prod_{i=1}^{L} \exp\left(-\frac{\beta U n_i(n_i - 1)}{2}\right) \delta\left(\sum_{i=1}^{L} n_i - N\right), \label{meas}
\end{align}
where $\mathcal{Z}$ is the canonical partition function.

For small bias $g$, the system remains close to equilibrium, allowing the use of the unperturbed (equilibrium) measure defined in Eq.~\eqref{meas} to compute the linear response. The conductivity $\sigma$ is defined as
\begin{align}
    \sigma = \lim_{g \to 0} \frac{j}{g},
\end{align}
where $j$ is the steady-state current. Within linear response, $\sigma$ can be estimated analytically by computing the expectation of the microscopic current in the equilibrium ensemble:
\begin{align}
    \sigma = \left\langle \min\left(1, \exp\left[-\beta U (n_{i+1} - n_i + 1)\right]\right) \theta(n_i) \right\rangle_{\mu},
\end{align}
where $\theta(n_i)$ is the Heaviside step function, and the average is taken with respect to the equilibrium measure $\mu$ defined in Eq.~\eqref{meas}.

To evaluate this analytically, we employed the grand canonical ensemble, adjusting the fugacity to fix the average density. The results were then compared with numerical simulations performed on a system of size $L = 150$, which was relaxed for $5L^2$ Monte Carlo steps {to achieve steady state}. The numerical estimate of the conductivity was obtained by measuring the slope of the current--bias ($j$--$g$) curve in the linear regime, using data up to $g = 0.10$.

\begin{figure}[h!]
    \centering
        \includegraphics[width=0.48\textwidth]{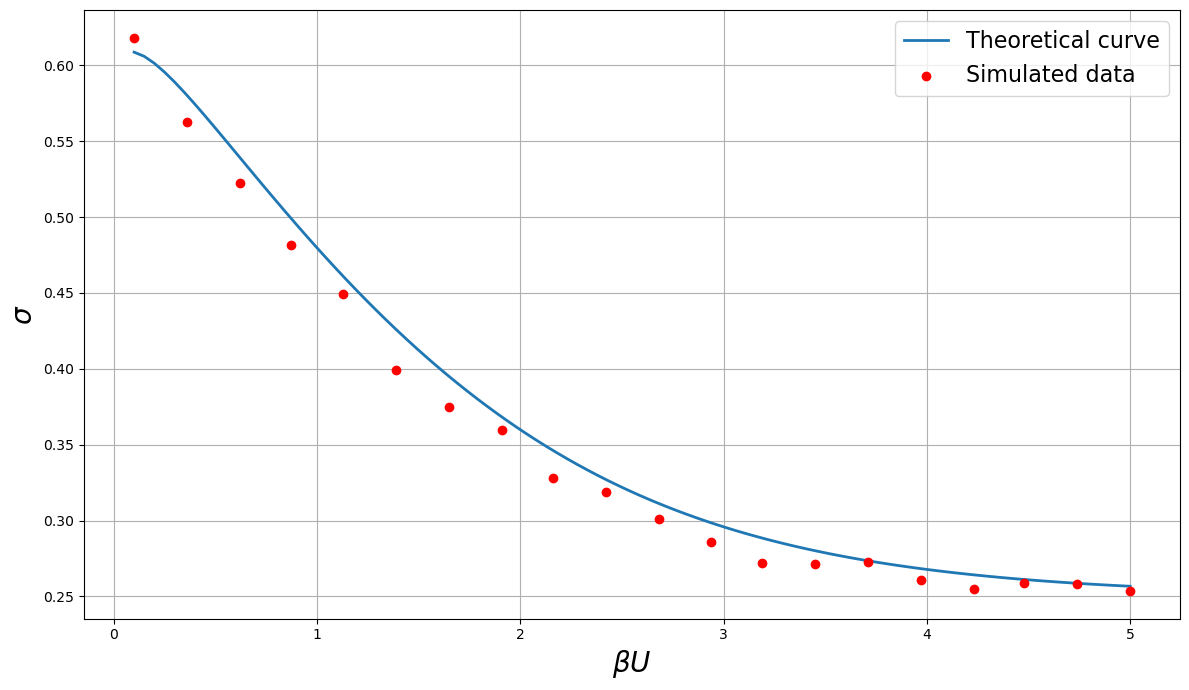}
        \caption{
      Comparison between analytic and simulated values of conductivity $\sigma$ for varying interaction strength $\beta U$, and fixed density $\rho=1.5$. Other simulation parameters, as for Fig.~\ref{fig:current_vs_g}. }
        \label{fig:lrt}
\end{figure}

We observed good agreement across a range of $\beta U$ values. The deviation from the analytic values arises from {finite size effects}.  A comparison of analytic and numerical results for the conductivity $\sigma$ is shown in Fig.~\ref{fig:lrt}. 

\renewcommand{\figurename}{FIG.}

%% file: sections/section04.tex
\section{Free Boundary Conditions} \label{sec:OBC}

We {now} consider a one-dimensional lattice with {free} boundary conditions, governed by the same dynamical rules described in Sec.~\ref{sec:model}. In the absence of on-site interactions, the system reduces to non-interacting particles undergoing biased hopping: a particle at site \( i \) hops to site \( i+1 \) with probability \( \frac{1+g}{2} \), and to site \( i-1 \) with probability \( \frac{1-g}{2} \). In general, the system can be interpreted as biased random walkers subject to an on-site repulsion.

We examine the steady-state density profiles obtained from the analytic solution and Monte Carlo simulations.
 Let \( n_k \) denote the occupation number at site \( k \), and define the average steady-state density as
\begin{equation}
    \rho_k = \langle n_k \rangle.
\end{equation}
These boundary conditions are also experimentally relevant, particularly in the classical limit of cold-atom systems, where Bose-Hubbard interactions in tilted optical lattices can be engineered~\cite{BHM,DipoleCond}.

The boundary conditions and system size \( L \) significantly influence the density profile. The bias \( g \) acts effectively as a constant external field that pushes particles toward site \( L \). Beyond a length scale, the energy gain from moving down the bias outweighs the repulsive energy cost, allowing particles to accumulate near the high-index end of the lattice.
\subsection{Analytic Solution}
In the case of free boundary conditions, there is no net current in the steady state. The system effectively consists of biased random walkers with on-site repulsion, evolving under free boundaries. In the grand canonical ensemble, the steady-state probability of a configuration $\mathbf{n}$ incorporates the effects of both the hopping bias and the interaction, and is given by
\begin{align}
    P(\mathbf{n}) = \prod_{k=1}^{L} \left( \frac{1}{\mathcal{Z}_k} \cdot {(z w_k)^{n_k}} \exp\left( -\frac{1}{2} \beta U n_k(n_k - 1) \right) \right),
\end{align}
where $z$ is the fugacity, and $w_k= \left( \frac{1+g}{1-g} \right)^k$ encodes the effect of the bias due to asymmetric hopping.

The local partition function at site $k$ is
\begin{align}
    \mathcal{Z}_k = \sum_{n=0}^{\infty} {(z w_k)^n}\exp\left( -\frac{1}{2} \beta U n(n - 1) \right).
\end{align}
{Thus the effective Hamiltonian which describes the equilibrium measure is}

{\begin{align}
       \mathcal{H}(n_1,n_2,\dots,n_L)=\sum_{k=0}^{L}\left(-Eakn_k+U\frac{n_k(n_k-1)}{2}\right).\label{hamil}
\end{align}}
{where:
\begin{align}
    \frac{1+g}{1-g}=e^{\beta E a}\label{repar},
\end{align}}
{$E$ is the effective external field strength, and $a$ is the lattice spacing.}
The average occupation at site $i$ is then given by
\begin{align}
    \langle n_i \rangle = \frac{1}{\mathcal{Z}_i} \sum_{n=0}^{\infty} n \cdot {(z w_i)^n} \exp\left( -\frac{1}{2} \beta U n(n - 1) \right).
\end{align}

\begin{figure}[h!]
    \centering
    \begin{subfigure}[b]{0.48\textwidth}
        \includegraphics[width=\textwidth]{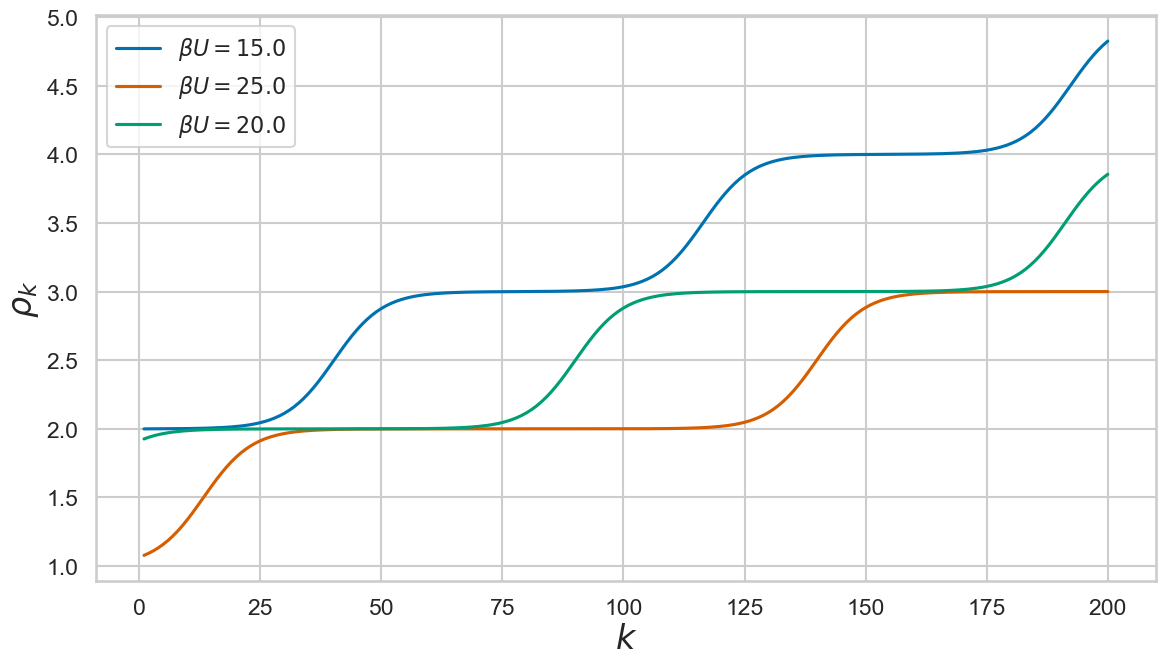}
        \caption{Average density $\rho_k$ versus lattice index $k$ at different values of $\beta U$, and overall density, with fixed $g=0.2,z=10^{10}$}.
        \label{fig:step}
    \end{subfigure}
    \hfill
    \begin{subfigure}[b]{0.48\textwidth}
        \includegraphics[width=\textwidth]{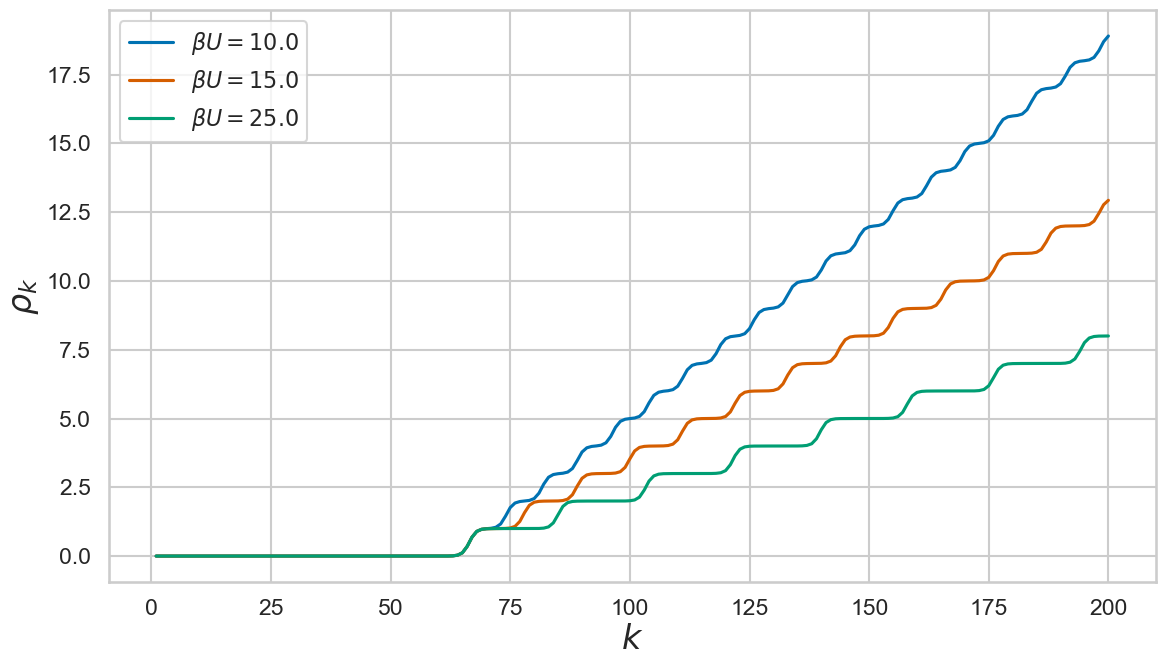}
        \caption{Average density $\rho_k$ versus lattice index $k$ at different values of $\beta U$, and overall density, with fixed $g=0.6,z=10^{-40}$.}
        \label{fig:linear}
    \end{subfigure}
    \caption{Steady state density profiles in open boundary conditions.}
    \label{fig:analobc}
\end{figure}

Depending on parameters, the resulting density profiles can exhibit two distinct features, for instance, a step-like increase in occupation or a depleted region near the boundary followed by a quasi-linear increase with small steps. These features are  {sensitive} to the system size, the parameter $w=\frac{1+g}{1-g}$, and the interaction strength $\beta U$. The analytic predictions from the above expressions show excellent agreement with numerical simulations, as demonstrated in Sec.~\ref{num}. The corresponding density profiles are displayed in Fig.~\ref{fig:analobc}.

\renewcommand{\figurename}{FIG.}
The emergence of density plateaus and the presence of a depleted region arise from the competition between the hopping bias and the interaction strength. However, the system size plays a crucial role: in sufficiently large systems, the effective biasing potential can outweigh the finite on-site repulsion, no matter how large. These profiles also highlight the pronounced impact of boundary conditions on the steady-state structure.
\subsection{Numerical Validation}\label{num}
The analytic results obtained above are well validated both qualitatively and quantitatively through Monte-Carlo simulations.
\subsubsection{Large \( L \) Limit}

In the limit of large \( L \), we observe a marked depletion of particles at low-index sites. A macroscopic region of the lattice may remain essentially unoccupied. A similar effect occurs as \( g \to 1 \), particularly when \( \beta U \) is small.

\begin{figure}[h!]
    \centering
    \includegraphics[width=0.48\textwidth]{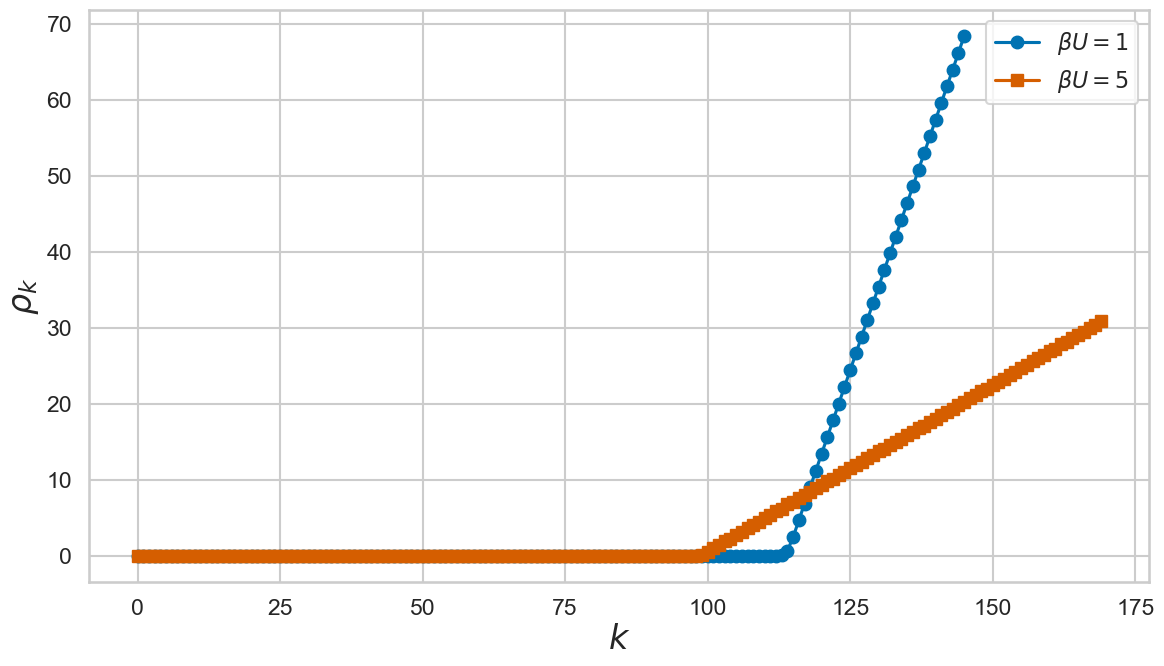}
    \caption{Average occupancy \( \rho_k \) as a function of site index \( k \), for various values of \( \beta U \) at fixed \( g = 0.8 \).Other simulation parameters: system size $L = 200$; $10^6$ Monte Carlo steps; relaxation time $5L^2$; global density=$5.5$.}
    \label{fig:rho2}
\end{figure}

As shown in Fig.~\ref{fig:rho2}, the biasing force becomes the dominant contribution at small \( \beta U \), leading to particle accumulation near the boundary. The density profile exhibits an approximately linear increase across a significant portion of the lattice. This behavior is governed by the interplay between the bias and interaction: when the bias-induced energy gain exceeds the interaction penalty, particles tend to cluster at the high-index sites.

\subsubsection{Finite System Size with Large \texorpdfstring{$\beta U$}{βU}}

For finite system sizes and large values of \( \beta U \), the repulsive interaction can overcome the energy gain induced by the external bias. In such regimes, we observe the formation of density plateaus—regions where particle occupancy saturates—reflecting the suppression of further accumulation due to interaction energy costs.

\renewcommand{\figurename}{FIG.}
\begin{figure}[h!]
    \centering
    \includegraphics[width=0.48\textwidth]{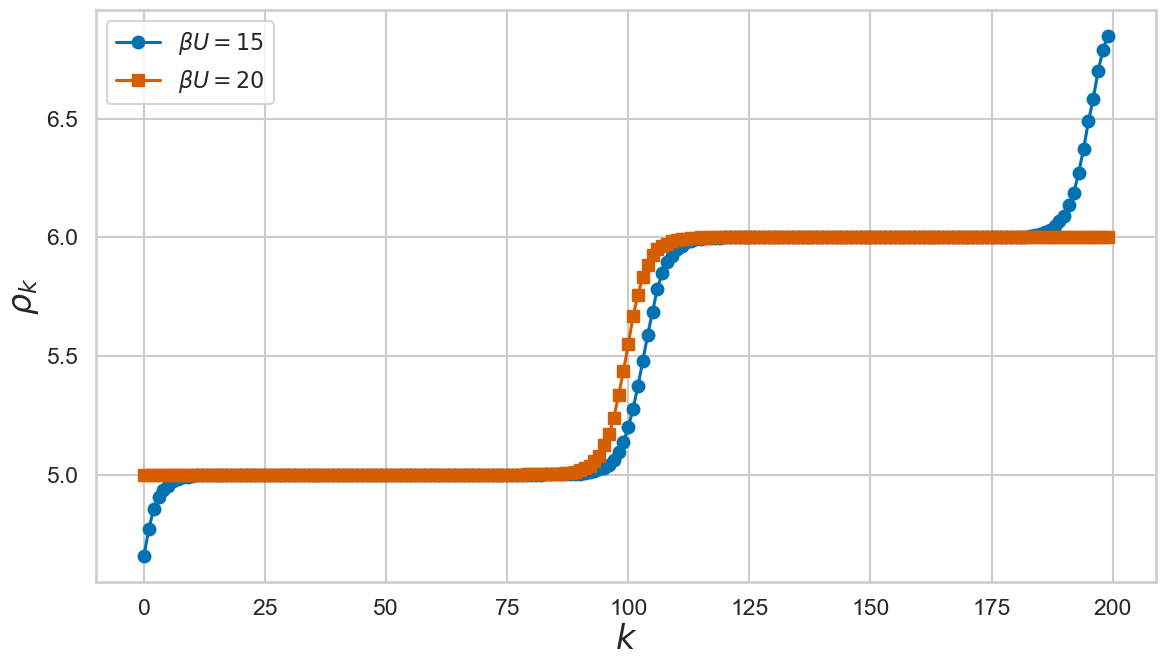}
    \caption{Average occupancy \( \rho_k \) as a function of site index \( k \), for various values of \( \beta U \) at fixed bias \( g = 0.2 \). Other simulation parameters, as for Fig.~\ref{fig:rho2}.}
    \label{fig:bu}
\end{figure}

As shown in Fig.~\ref{fig:bu}, for sufficiently large \( \beta U \) and small bias \( g \), the density profile exhibits extended flat regions indicative of local occupancy saturation. This behavior signals the dominance of interaction energy over the bias-induced drift, preventing significant particle accumulation even near the biased edge of the lattice.

%% file: sections/section06.tex
\section{Conclusion} \label{sec:conc}

In this work, we investigated a one-dimensional stochastic lattice model with Bose–Hubbard-type on-site interactions and asymmetric hopping dynamics, bridging key features of ASEP, ZRP, and the classical Bose–Hubbard model. Our analysis, combining exact limits and Monte Carlo simulations, revealed rich steady-state behavior under both periodic and open boundary conditions.

For periodic systems, we observed that nearest-neighbor correlations vary non-monotonically with the interaction strength. This trend stems from a competition between bias and repulsion. The steady-state current undergoes a crossover, appearing ZRP-like in the weak-interaction limit and ASEP-like when interactions are strong. At large interaction strengths, both the current and the variance fluctuations change periodically with particle density, as a consequence of the emergence of an ASEP which describes remnant particles moving on a quiescent background of stacks of immobile particles.

With {free} boundary conditions—relevant for cold atom systems in tilted lattices—the interplay of repulsion and drive produces {interesting} density profiles, ranging from uniform plateaus to steep gradients following regions of depletion. These features underscore the importance of boundary effects in shaping steady states.

Our study illustrates how controlled stochastic dynamics with tunable interactions and bias can serve as a minimal framework for exploring complex transport phenomena in driven many-body systems.

 Several open questions remain. First, a complete analytical solution for the steady-state distribution at finite interaction strength {and bias,} remains elusive due to the non-factorizable nature of the dynamics. Second, time-dependent properties such as relaxation dynamics, spectral gaps, and fluctuation theorems merit investigation. Lastly extending this study to quenched disorder could shed light on localization phenomena due to the interplay between interactions and bias. This is illustrated by a study of particles with Bose-Hubbard interactions on a random comb~\cite{RC}. 

%% file: sections/acknowledgements.tex
\section*{Acknowledgements} \label{sec:acknowledgements}
We thank Prof. Arti Garg for drawing our attention to the similarity between our model and Bose–Hubbard type interactions, and for insightful discussions on this topic. S.M. thanks TIFR Hyderabad for supporting this work during the Visiting Students Research Programme (VSRP) and subsequent visits as a visiting student. M.B. acknowledges the support of the Indian National Science Academy (INSA). We also acknowledge the support of the Department of Atomic Energy, Government of India, under Project Identification No. RTI4007.